\documentclass[letterpaper,preprint,superscriptaddress,floatfix,showpacs]{revtex4}
\usepackage[latin1]{inputenc}
\usepackage{bm}
\usepackage{multirow,amssymb,amsbsy,amsmath}
\usepackage{graphicx}
\usepackage{verbatim}
\makeatletter
\usepackage{pifont}
\makeatother

\begin{document}

\title{Realization of reliable solid-state quantum memory for photonic polarization-qubit}

\author{Zong-Quan Zhou}
\affiliation{Key Laboratory of Quantum Information, University of
Science and Technology of China, CAS, Hefei, 230026, China}

\author{Wei-Bin Lin}
\affiliation{Key Laboratory of Quantum Information, University of
Science and Technology of China, CAS, Hefei, 230026, China}

\author{Ming Yang}
\affiliation{Key Laboratory of Quantum Information, University of
Science and Technology of China, CAS, Hefei, 230026, China}

\author{Chuan-Feng Li$\footnote{email:cfli@ustc.edu.cn}$}
\affiliation{Key Laboratory of Quantum Information, University of
Science and Technology of China, CAS, Hefei, 230026, China}

\author{Guang-Can Guo}
\affiliation{Key Laboratory of Quantum Information, University of
Science and Technology of China, CAS, Hefei, 230026, China}

\date{\today}

\begin{abstract}
{Faithfully storing an unknown quantum light state is essential to
advanced quantum communication and distributed quantum computation applications. The required quantum memory must have high fidelity to
improve the performance of a quantum network. Here we report the reversible transfer of photonic
polarization states into collective atomic excitation in a compact solid-state device.
The quantum memory is based on an atomic frequency comb (AFC) in rare-earth ion doped crystals. We obtain up to $0.998$ process
fidelity for the storage and retrieval process of single-photon-level coherent pulse. This reliable quantum
memory is a crucial step toward quantum networks based on solid-state
devices.}
\end{abstract}

\maketitle

Quantum communication channels are naturally carried by photons. To realize a
quantum network, we must coherently transfer quantum information between the
stationary qubits and the flying photon qubits \cite{Tittle09,Duan10}.
Despite some remarkable efforts
in quantum networks based on cold atoms \cite{Atom04,Atom08,Atom09,Atom11}, single atom in a cavity \cite{SingleAtom11}
Bose-Einstein condensate \cite{BEC11}, atomic vapors \cite{Vapor09,Vapor11} and
trapped ions \cite{Duan10,Ion09}, there are strong motivations for using
more practical systems, e.g. solid-state devices. Rare-earth (RE) ion-doped
solids provide a particular electronic structure that can be seen as a frozen
gas of atoms, they have excellent coherence properties for optical and spin
transitions. Moreover, they have already shown excellent capability to store
light for extended periods \cite{time05} with high efficiency \cite{eff10} and
a large bandwidth \cite{bw1,bw2,bw3}. Recent achievements include the storage
of photonic time-bin entanglement generated through spontaneous parametric down
conversion (SPDC) \cite{bw3,timebin}.

Logical qubits with photons can be encoded in several ways, for example via
polarization, time-bin, path, phase or photon-number encodings. The
polarization degree of freedom is particularly useful because it allows a
single-photon qubit to be transmitted in a single spatial and temporal mode.
Polarized photons are more easily transportable qubits and are more robust
against decoherence. Note that many quantum light sources generate entangled
photons with information encoded in their polarization degree, such as the
eight-photon entangled state generated by SPDC \cite{ghz} and the on-demand
entangled photon pairs generated by biexciton decay in quantum dots \cite{qd}.
Realizing the quantum interface between quantum memory and other quantum light
source will also require the capability of the quantum memory to store the
polarization information for light. Therefore, the ability to store
polarization information with high fidelity is of particular interests. One
approach to realizing a polarization-qubit memory is to first transform the
polarization encoding into path encoding and then store it into two distant
ensembles. This approach has been implemented in both atomic ensembles
\cite{Atom04,Atom08,Atom11} and atomic vapors \cite{vapor2}. The abovementioned
realizations suffer from several practical drawbacks due to the extra step of
spatially splitting the input state. Phase locking of the two optical paths
\cite{Atom11} may be required, and imperfect spatial mode matching may limit
the memory fidelity. Another approach is using uniformly absorbed samples for
two orthogonal polarizations, which requires no spatially splitting of the
input photon. This approach has been implemented with two spatially overlapped
atomic ensembles \cite{Atom09}, BEC \cite{BEC11} and with a single-atom in a
cavity \cite{SingleAtom11}. Up to now, the best fidelity performance achieved
with single-photon-level input is 0.95 \cite{BEC11}. All of these experiments
are based on the Raman process or electromagnetically induced transparency
(EIT). These processes require a strong control or read light pulse during the
memory sequence, which introduces unavoidable noise into the retrieval signal.

AFC is an alternative protocol for realizing quantum memory without a strong
control light during the memory sequence. AFC quantum memory with RE doped
solids has shown an excellent capability to store quantum light
\cite{AFC08,bw1,bw2,bw3,timebin,AFCeff,AFCspin}. Because of the strongly
polarization-dependent absorption of ions in crystal, however, all previous
experiments have been conducted with a single pre-defined polarization. Here,
we use an alternative approach to realize AFC quantum memory for
polarization-encoded single photons. Two pieces of crystals are used to absorb
the orthogonal polarized components that are in the same optical path but at
different sites.

The hardware of our quantum memory, which is shown in Fig. 1(a), is composed of
two pieces of Nd$^{3+}$:YVO$_4$ crystals (doping level 10 ppm) sandwiching a
half-wave plate (HWP). The sizes of the two crystals are nearly equal with 1.4$\pm0.01$
$mm$ length along the a-axis. The $^{4}I_{9/2}\rightarrow{ }^4F_{3/2}$
transition of Nd$^{3+}$ at about 879.705 nm in the Nd$^{3+}$:YVO$_4$ crystal
shows strong absorption of $H$ polarized photons and little absorption of $V$
polarized photons \cite{SHB}. Here, $H (V)$ denotes horizontal (vertical),
which is defined to be parallel (perpendicular) to the crystal's c-axis. This
prevents a single piece of Nd$^{3+}$:YVO$_4$ crystal from functioning as a
polarization-qubit memory because most of the $V$ polarized photons will pass
through the crystal without being absorbed. By using a $45^\circ$ HWP,
sandwiched between two parallel Nd$^{3+}$:YVO$_4$ crystals, the sample shows
near equal absorption depth for $H$, $V$ and arbitrary polarized photons.

\begin{figure}[tb]
\centering
\includegraphics[width=0.8\textwidth]{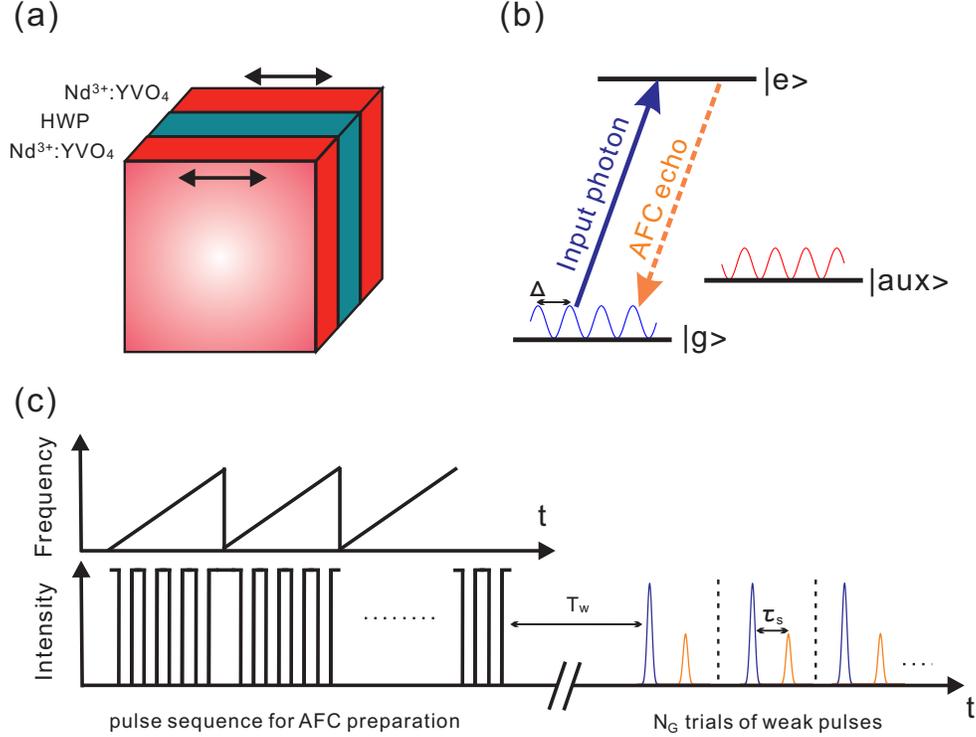}
\caption{\label{Fig:1} (Color online) (a). Illustration of the sample used as the memory
hardware for arbitrary polarizations. The arrows represent the c-axis of the
crystals. Details about the sample are given in the Supplementary Information. (b). The near 880 nm input photons are absorbed at the
$^{4}I_{9/2}\rightarrow{ }^4F_{3/2}$ transition of Nd$^{3+}$. This transition
is strongly $H$ polarized. After a programmable time, the photon will be
collectively emitted in a desirable spatial mode. (c). The preparation and
storage sequence used in the experiment. }
\end{figure}

The AFC protocol requires a pumping procedure to tailor the absorption profile
of an inhomogeneously broadened solid state atomic medium with a series of
periodic and narrow absorbing peaks separated by $\Delta$ (see Fig. 1). The
single input photon is then collectively absorbed by all of the atoms in the
comb. The atomic state can be represented by, \cite{AFC09},
\begin{equation} \label{eq1}
|1\rangle_A=\sum_jc_je^{i\delta_jt}e^{-ikz_j}|g_1\cdot\cdot\cdot e_j
\cdot\cdot\cdot g_N\rangle.
\end{equation}
Here $N$ is the total number of atoms in the comb; $|g_j\rangle$ and
$|e_j\rangle$ represent the ground and excited states, respectively, of atom
$j$; $z_j$ is the position of atom $j$; $k$ is the wavenumber of the input
field; $\delta_j$ is the detuning of the atom frequency and the amplitudes
$c_j$ depend on the frequency and on the position of atom $j$. The atoms at
different frequencies will dephase after absorption, but because of the
periodic structure of AFC, a rephasing occurs after a time
$\tau_s=2\pi/\Delta$. The photon is re-emitted in the forward direction as a
result of a collective interference between all of the atoms that are in phase.
The collective optical excitation can be transferred into a long-lived ground
state to achieve a longer storage time and an on-demand readout \cite{AFCspin}.

The AFC preparation pulse sequence are shown in Fig. 1(c), in which the pump light's frequency is swept over 100 MHz in 100 $us$ and each frequency step has been assigned a specific amplitude to give an comb structure. The
pumping sequences are repeated 100 times to achieve an optimal AFC structure. The duty ratio in each cycle is carefully adjusted to take into account the power broadening effect. An experimental AFC structure with a periodicity of 20 MHz and bandwidth of 100 MHz is given in the Supplementary Information. The AFC with large bandwidth also can be realized with temporal pulse pairs and frequency shifted pumping technique \cite{bw1}. We note that the AFC created directly in the frequency domain shows a better performance of memory efficiency. During the storage cycle, $N_g$ trials of single-photon-level coherent pulses
are sent into the sample, and the AFC echoes are emitted at a time
$2\pi/\Delta$ that is programmed in the pumping procedure. To avoid the
fluorescence noise caused by the classical pumping light, the memory cycle
begins after waiting for time $T_w=1.2$ $ms$ after the preparation
cycle is completed. The complete pump and probe cycles are repeated at a frequency of
40 Hz.

\begin{figure}[tb]
\centering
\includegraphics[width=0.8\textwidth]{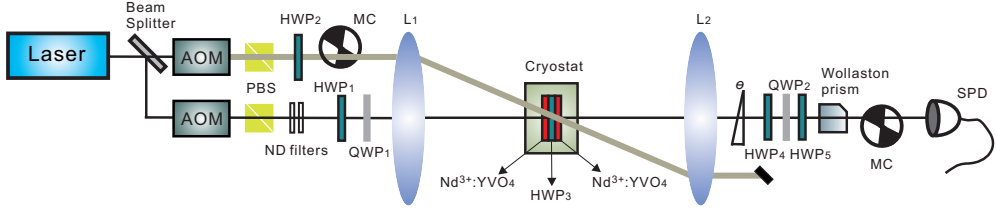}
\caption{\label{Fig:2} (Color online) The experimental setup for quantum storage of a photonic
polarization-qubit in solids. The upper acousto-optic modulators (AOM) produces
pump light for the AFC preparation. The lower AOM produces a weak probe light
for the storage sequence. The light's polarizations are initialized by the
polarization beam splitters (PBS). The probe light's polarizations are
controlled by the half-wave plate (HWP$_1$) and quarter-wave plate (QWP$_1$).
The samples are placed in a cryostat at a temperature of 1.5 K. Beyond the
sample, a phase plate ($\theta$) and HWP$_4$ corrects any polarization rotation
caused by the sample. The probe light's polarizations are then analyzed with
QWP$_2$, HWP$_5$ and a Wollaston prism. The two MCs are used to protect the
single-photon detector (SPD) from classical pump pulses. }
\end{figure}
The experimental setup is shown in Fig. 2. The laser source is a cw Ti:sapphire
laser (M Squared, Solstis), and the laser's wavelength is monitored with a
wavelength meter (Bristol, 621A). The pump light is generated with a AOM with a center frequency of 260 MHz
(Brimrose) in a double-pass configuration. To protect the single-photon
detectors during the preparation procedure, a mechanical chopper (MC) is placed
in the pumping optical path. To achieve a uniform memory efficiency for
arbitrary polarized photons, the pumping light should be polarized close to
$H+V$. The $H$ polarized light will be mostly absorbed by the first crystal,
and the $V$ polarized light will be mostly absorbed by the second crystal,
after the polarization rotation by the HWP$_3$. Carefully adjusting the HWP$_2$
angle and the pump power can optimize the storage efficiency and balance the
efficiency of the H and V components. The photons to be stored are generated by
another 260 MHz AOM in double-pass configuration. The AOMs are controlled by an
arbitrary function generator (Tektronix, AFG3252) and the direct digital synthesizers (Isomet). The HWP$_1$ and QWP$_1$
prepares input photons with arbitrary polarizations. The input photons are
decreased to single-photon level by the neutral density (ND) filters. The
storage sequence is repeated 1600 times at a frequency of 1 MHz.

Note that the beam splitters and single mode fiber may degrade the fidelity of
the input photons. To achieve a optimum polarization storage performance, we
use a setup that differs from those of previous experiments, in which the pump
light and probe light have been in counterpropagation or copropagation
configurations. The pump light is only overlapped with the probe light at the
sample. The pump light and probe light are focused with the same lens L$_1$ (focal
length: 250 $mm$). This setup achieves a small angle ($\sim15$ $mrad$) between
pump and probe lights. The probe light focuses to a diameter of about 100 $um$,
while the pump light is collimated to produces a much larger diameter on the
sample. The sample is placed in a cryostat (Oxford Instruments, SpectromagPT)
at a temperature of 1.5 K and with a magnetic field of 0.3 T in the horizontal
direction. The two parallel Nd$^{3+}$:YVO$_4$ crystals' c axes are placed in
the horizontal direction. The HWP$_3$ at 45$^\circ$ can exchange $H$ and $V$
polarizations. Because each crystal only strongly absorbs $H$ polarized light,
the $H$ polarized input photon is stored in the first crystal, and the
$V$ polarized light is stored in the second crystal. Another lens L$_2$ collects the photons outside of the cryostat. To rotate the polarization
back to the input state, another HWP (HWP$_4$) is placed at 45$^\circ$ outside
of the cryostat. Because of the small difference in the lengths of the two
crystals, a phase plate is inserted in the optical path to compensate for the
small phase shift between $H$ and $V$ polarized photons. The QWP$_2$, HWP$_5$
and Wollaston Prism together choose the polarization of the single photon
detections. Another MC is used to protect the single-photon detector (PerkinElmer, SPCM AQRH-15) from
the classical pump light. The signal from the SPD is sent to the time-interval
analyzed (TIA) and time-correlated single photon counting (TCSPC) system
(Picoquant, Hydraharp 400).

Fig. 3(a) shows an example of weak coherent pulses (the average photon number
is 0.8 photons per pulse) with a duration of 25 ns and a polarization of $H+V$
that are collectively mapped onto the sample. A strong echo is emitted after a
preprogrammed storage time of 200 ns. The measured storage and retrieval
efficiency is about $6.9\%$. The mean input photon number is determined by measuring the
detection probability per pulse with the laser 20 GHz off resonance, and take into account the detection efficiency ($\sim 0.4$) and the transmission from the
sample to the detector ($\sim 0.6$).

\begin{figure}[tb]
\centering
\includegraphics[width=0.8\textwidth]{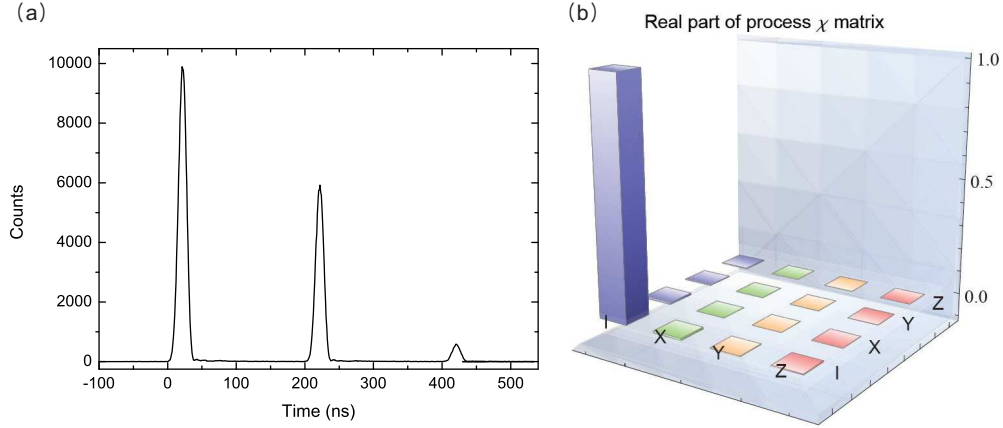}
\caption{\label{Fig:3}(Color online) (a). With an AFC prepared with a periodicity of 5 MHz,
the $H+V$ polarized single-photon pulses are collectively re-emitted after a
200 ns storage time in the sample. The second echo are emitted after a storage time of 400 ns. (b). The real part numbers in the process
matrix $\chi$ as obtained from a quantum process tomography. All of the
imaginary numbers are close to zero, with the largest amplitude being 0.018.}
\end{figure}

To characterize the polarization storage performance of the memory, we perform
complete quantum process tomography (QPT) \cite{QPT,QPT2} on the system (see
Supplementary Information). The tomography result of the process matrix $\chi$ with a memory time
of 200 ns are shown in Fig. 3(b). The results give a process fidelity of
$0.998\pm0.003$ for an average input photon number of 0.8 photons per pulse.
Note that no dark counts of SPD are corrected
through our experiment. Our results are far beyond the 2/3 bound, which is the
maximum average fidelity that can be achieved with a classical memory
\cite{bound}. Controlled reversible inhomogeneous broadening (CRIB)
\cite{Vapor09,Vapor11,eff10} is a protocol that is similar to AFC. CRIB uses an
external field to rephase the atoms and thus eliminates any classical light
during the storage cycle; therefor, it should achieve a similar fidelity
performance. Our scheme is applicable to the CRIB protocol.

\begin{figure}[tb]
\centering
\includegraphics[width=0.5\textwidth]{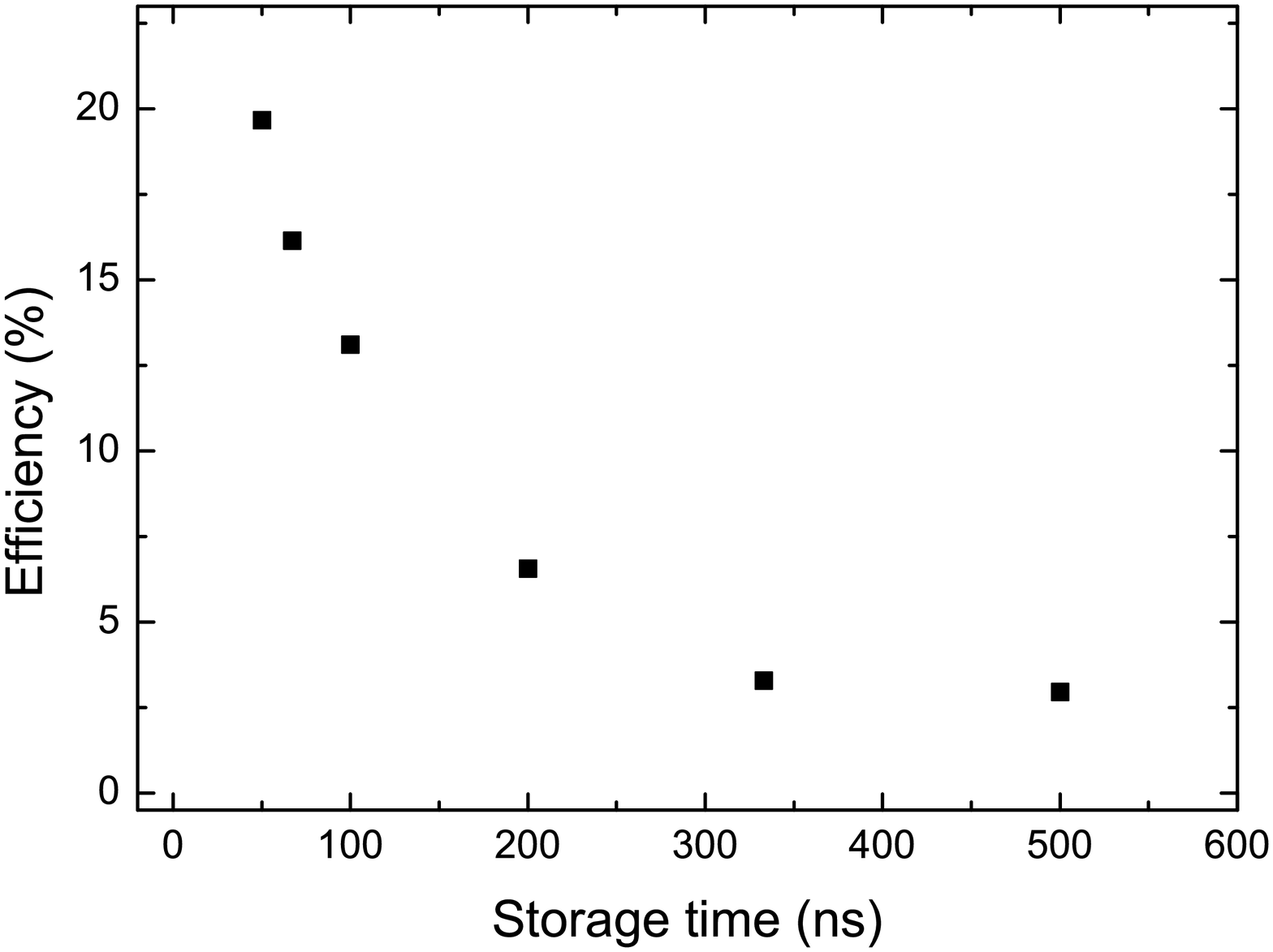}
\caption{\label{Fig:4} The storage efficiency as a function of storage time.}
\end{figure}
By carefully designing the optical setup, we have eliminated most of the noise
from the setup or the environment. The remaining imperfections are mostly
introduced by the statistical photon drift and detector noise. We have also
characterized our memory by measuring the read-write fidelity as a function of
the storage time. The process fidelity shows little dependence on the storage
time. For a longer storage time (500 ns), the fidelity decreases to
$0.984\pm0.004$. Because the storage efficiency decreases quickly (as shown in
Fig. 4), the signal to noise ratio decreases. This factor account for most of
the degradation. In general, the echo-type memory based on collective
interference is not sensitive to decoherence, although the efficiency degrades
significantly. This phenomenon is elucidated by results showing that most of
the atoms that are decohered do not contribute to the collective interference,
as has been observed in classical photon-echo memory \cite{decoherence}. Our
low-noise scheme should enable accurate experimental investigations into the
influence of decoherence on memory fidelity.

We have demonstrated an ultra-reliable quantum memory for photonic
polarization-qubit in solids. By performing quantum process tomography, we
measure the process fidelity for a 200 ns storage time with single photon level
input pulses at $0.998\pm0.003$. This excellent fidelity performance should
make the quantum memory suitable for quantum error-correction applications in
large-scale quantum computation \cite{QEC10,qc}. Moreover, the absence of ions'
motion in the solid state means that even complex spatial structures can be
generated by light and stored. Our results should produce solid-state devices
that are capable of functioning as a hyper-quantum memory for light's
polarization, temporal, and spatial information.

We note that related results have been obtained by two other groups
\cite{ref1,ref2}.

This work was supported by the National Basic Research Program (2011CB921200), National Natural
Science Foundation of China (Grant Nos. 60921091 and 10874162).

\section*{Supplementary Information}

\subsection*{Sample} The sample is composed of two pieces of Nd$^{3+}$:YVO$_4$
crystals (doping level 10 ppm), sandwiching a half-wave plate (HWP). The sizes
of the two crystals are nearly equal, with 10*10 $mm$ apertures and 1.4 $mm$
lengths along the a-axis. The absorption of a single piece of Nd$^{3+}$:YVO$_4$
crystal shows strong polarization dependence, with absorption coefficients of
$\alpha$ of 4.6 $/mm$ for $H$ polarized light and $\alpha$ of 0.7 $/mm$ for $V$
polarized light. To improve the accuracy of the crystal alignment and the HWP,
the HWP is designed as a 10*10 $mm$ square with an optical axis at the
$45^\circ$ diagonal line of the square. This configuration can exchange the $H$
and $V$ polarizations with an extinction ratio of better than $2000:1$. The
complete experimental setup gives an extinction ratio of about $900:1$ for the
$H+V$ and $H-V$ polarizations.

\subsection*{Quantum process tomography} The quantum process can be represented by
the process matrix $\chi$. The $\chi$ matrix can be expressed on the basis of
$\sigma_m$ and maps an input density matrix $\rho_{in}$ onto the corresponding
output state $\rho_{out}$ via
\begin{equation} \label{eq1}
\rho_{out}=\sum_{m,n=0}^{3} \chi_{mn}\sigma_m\rho_{in}\sigma_{n}^{\dag}.
\end{equation}
We chose is $[I, X, Y, Z]$ as the basis for $\sigma_i$, where I represents the
identity operation and X, Y and Z represent the three Pauli operators. Here,
$\dag$ denotes the adjoint operator. The $\chi$ matrix completely and uniquely
describes the process and can be reconstructed by experimental tomographic
measurements. We prepare and measure four polarization states ($H$, $V$,
$R=H+iV$, and $D=H+V$) and perform quantum state tomography for the output
state generated by each input pulse. In the experiment, the physical matrix
$\chi$ is estimated by the maximum-likelihood procedure, which is
shown in Fig. 3(b). The fidelity of the experimental result is about 0.998, as
calculated by $(Tr\sqrt{\chi}\chi_{ideal}\sqrt{\chi})^2$ with $\chi_{ideal}=1$.

\subsection*{AFC structure}
In our experiment, an optimized efficiency of about $20.8\%$ at 50 ns storage time is obtained. We measure the AFC structure by sending long probe pulse and measuring their transmission. An example AFC with a periodicity of 20 MHz is shown in the Fig. S1. Refer to the AFC structure, our efficiency promotion is mainly caused by the large available optical depth ($\sim6.2$) and small background absorption ($\sim0.5$). The experimental efficiency fits well the theoretical estimation.

\begin{figure}[tb]
\centering
\includegraphics[width=0.5\textwidth]{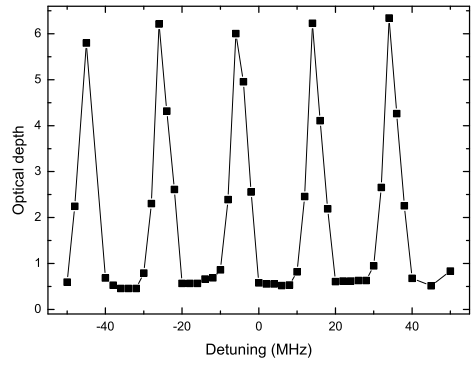}
\caption{\label{Fig:S1} An AFC prepared with a periodicity of 20 MHz.}
\end{figure}




\end{document}